\documentstyle[amssymb,aps]{revtex}
\tightenlines
\begin{document}
\title{Collective flux creep: beyond the logarithmic solution}
\author{L. Burlachkov, D. Giller, and R. Prozorov}
\address{Institute of Superconductivity, Department of Physics, Bar-Ilan University,\\
Ramat-Gan 52900, Israel}
\date{February 6, 1998}
\maketitle

\begin{abstract}
Numerical studies of the flux creep in superconductors show that the
distribution of the magnetic field at any stage of the creep process can be
well described by the condition of spatial constancy of the activation
energy $U$ independently on the particular dependence of $U$ on the field $B$
and current $j$. This results from a self-organization of the creep process
in the undercritical state $j<j_{c}$ related to a strong non-linearity of
the flux motion. Using the spatial constancy of $U$, one can find the field
profiles $B(x)$, formulate a semi-analytical approach to the creep problem
and generalize the logarithmic solution for flux creep, obtained for $U=U(j)$%
, to the case of essential dependence of $U$ on $B$. This approach is useful
for the analysis of dynamic formation of an anomalous magnetization curve
(''fishtail''). We analyze the quality of the logarithmic and generalized
logarithmic approximations and show that the latter predicts a maximum in
the creep rate at short times, which has been observed experimentally. The
vortex annihilation lines (or the sample edge for the case of remanent state
relaxation), where $B=0$, cause instabilities (flux-flow regions) and modify
or even destroy the self-organization of flux creep in the whole sample.
\end{abstract}


\section{Introduction}

Since the discovery of the giant vortex creep \cite{YM} in high-temperature
superconductors (HTSC), it has become clear that the relaxation processes in
these compounds may be very rapid compared to usual low-temperature
superconductors. The magnetization current $j$ and, in turn, the magnetic
moment $M$, which is approximately proportional to $j$ in most cases, drop
considerably during the usual experimental time windows of a few hours (or
even less) down to small values $j\ll j_c$, in particular, at elevated
temperatures. Here $j_c$ is the critical current, which divides the regimes
of {\em flux creep} ($j<j_c$) characterized by the Boltzmann factor $\exp
(-U/kT)$, where $U(B,j,T)$ is the activation energy for flux creep, and the
non-activational {\em flux flow} ($j>j_c$) with $U=0$. Due to such a
pronounced relaxation, both $j$ and $M$ are determined in HTSC mostly by the
flux dynamics in contrast with conventional superconductors where relaxation
is usually very slow, so $j\cong j_c$ for any accessible time windows. This
has given rise to an extensive study, both theoretical and experimental, of
magnetic relaxation and vortex dynamics in HTSC (see \cite
{Blatter-rev,YSM-rev,Brandt-rev} as reviews). Most of them are based on the
logarithmic solution \cite{GeshLar}: 
\begin{equation}
U\simeq kT\ln \left( 1+\frac t{t_0}\right) ,  \label{eq-Ulog}
\end{equation}
where $t_0$ is the logarithmic time scale for flux creep. We will discuss
Eq.~(\ref{eq-Ulog}) in detail in Section III.

A closely related problem is the so-called ''fishtail'' effect, i.e., the
anomalous increase of $M$ as a function of $H$ \cite{fisht}, or the increase
of locally measured $j$ as a function of $B$ \cite{Abulafia2,Giller}. This
effect serves as a test for different models of flux pinning and creep. The
crucial question is whether the non-monotonous behavior of $j$ results from
the same feature in $j_c$ (''static fishtail''), or arises from a faster
relaxation of $j$ at small $B,$ whereas $j_c$ is a monotonously decreasing
function of $B$ and itself shows no anomaly (''dynamic fishtail''). The
latter possibility implies that the fishtail effect should disappear at
shorter time windows or lower temperatures where the effect of relaxation is
negligible and $j\cong j_c$ \cite{Bont}.

After the instantaneous switching on of the external field $H$, the
flux-flow process develops towards establishing a nearly critical profile $%
j\cong j_c(B)$, where the Lorentz force $\left( j_c\times \phi _0\right) /c$
is compensated by the pinning force ($\phi _0$ is the flux quantum and $c$
is the velocity of light) all over the sample. Usually the duration $\tau
_{flow}$ of flux flow does not exceed a few milliseconds (see Ref.~\cite{FGV}
and references therein). As $j$ drops below $j_c$, the slow process of flux
creep starts. The creep rate is mostly determined by the Boltzmann factor $%
\exp (-U/kT)$, where $U=U(B,j)$. Of course, $U$ depends also on temperature,
but the creep experiments are usually conducted at constant temperature. In
many cases the dependence of $U$ on $B$ can be neglected, which implies a
crucial simplification for the theoretical description of flux creep. For
instance, in an infinite slab of width $2d$ ($-d<x<d$) in the parallel
external field $H$, the variation of the magnetic induction $\delta
B=B(d)-B(0)$ do not exceed $H^{*}$, where $H^{*}=(4\pi /c)j_cd$ is the full
penetration field \cite{Bean} ($j_c$ is considered to be field-independent).
For $H\gg H^{*}$ one can neglect $\delta B\lesssim H^{*}$. If both $j_c$ and 
$d$ are sufficiently small, say, $j_c\simeq 10^4$ ${\rm A/cm}^2$ and $%
d\simeq 0.1$ ${\rm mm}$, then $H^{*}$ is of order hundreds of Gauss. In this
case the above condition is easily fulfilled for most $H$, and the
activation energy appears to be field-independent: $U(j,B)\cong U(j,H)$.
Then the field profiles $B(x)$ are almost straight \cite{FGV,Beek-short},
i.e., $j\cong const$ throughout the sample at all creep stages.

However, in larger samples ($d\gtrsim 1\ {\rm mm}$) with strong pinning ($%
j_c>10^5~{\rm A/cm}^2$), one gets $\delta B\simeq H^{*}>1\ {\rm Tesla}$,
which implies that the spatial variation of $B$ may not be small compared to 
$H$. Of course this estimation may not be valid at high temperatures where $%
j_c$ drops. But for large and not too clean samples well below $T_c$ the
dependence of $U$ on $B$ is essential, and the field profiles are not
straight. On the other hand, due to very fast relaxation $B$ can vary by
orders of magnitude during experimental time windows of order of hours (see,
for instance, \cite{Abulafia2}), which also requires taking the $U(B)$%
-dependence into account for the consistent description of the relaxation
process, especially in the dynamic models of the fishtail formation.

The goal of this paper is to study the relaxation process for various
dependencies of $U$ on $j$ and $B$. In Section II we show that the field
profile at any stage of the relaxation process can be described by the
condition of spatial constancy of the activation energy: 
\begin{equation}
U(x)\cong const  \label{eq-const}
\end{equation}
throughout the whole sample, where $const$ depends on time only. Since $%
U=U(B,j)$, Eq.~(\ref{eq-const}) provides an implicit relationship between $B$
and $j$, manifesting a condition of {\em self-organization} of flux motion
in the undercritical regime $j<j_c$. In other words, according to Eq.~(\ref
{eq-const}) the field profiles form a one-parameter family $B_U(x)$. The
problem to be solved in order to describe the flux creep is to find these
profiles together with the dependence of $U$ on time. Note that this case
differs from the self-organized criticality (see the pioneering work \cite
{Bak} and further applications to the superconductors in a critical state 
\cite{SOC-HTSC}), where the peculiarities of the flux motion (avalanches,
critical exponents, etc.) are considered in the vicinity of the critical
state $j\cong j_c$ (i.e., at $U\cong 0$). The condition $j\cong const$ found
in previous studies \cite{FGV,Beek-short} is obviously just a particular
case of Eq.~(\ref{eq-const}) provided $U$ is independent of $B$. In Section
III we analyze numerically and, using Eq.~(\ref{eq-const}),
semi-analytically the flux creep for various $U(B,j)$-dependencies,
particularly for the most general collective creep behavior $U\propto
B^\alpha j^{-\mu }$. We show that at short, but experimentally available
time scales the creep process differs significantly from the logarithmic
solution (see Eq.~(\ref{eq-Ulog})) and shows a maximum in the relaxation
rate, $dU/d\ln t$, in accordance with the experimental data. The
semi-analytical solution provides a good approximation to the exact
(numerical) one at all time scales. In Section IV we apply these ideas to
the problem of the anomalous magnetization curve (fishtail) and show how the
dynamic development of the anomaly in $j$ (or the same, in $M\propto j$) can
be described semi-analytically. In Section V we study the effect of
so-called ''annihilation lines'' in infinitely long samples, where $B$
changes sign and vortices with opposite directions annihilate each other, on
the self-organization of the flux motion. A particular case of such a line
is the edge of the infinitely long sample in the remanent state. The vortex
velocity $v$ shows a peculiarity (divergence) at such a line, resulting in
the appearance of flux-flow regions in the vicinity of the annihilation
lines. We show that these peculiarities affect deeply the self-organization
of creep and the condition of spatial constancy for $U$ (see Eq.~(\ref
{eq-const})) is modified or even destroyed in the whole sample (and not only
in the vicinity of the annihilation lines).

\section{Spatial constancy of $U$}

Consider an infinite slab of thickness $2d$ $(-d<x<d)$ with the magnetic
field $B(x)$ parallel to $z$-direction and the current $j$ flowing along $y$%
. The external field $H$ is switched on instantaneously when no vortices are
present in the sample, which corresponds to zero-field-cooled experiments.
We consider $H\gg H_{c1}$, where $H_{c1}$ is the lower critical field, and
disregard the effects related to the latter.

The flux creep is described by the diffusion equation \cite{BLW}: 
\begin{equation}
\frac{\partial B}{\partial t}=-\frac{\partial D}{\partial x},
\label{eq-diff}
\end{equation}
where 
\begin{equation}
D=Bv={\cal A}\frac{\phi _0}{c\eta }Bj\exp (-U/kT)  \label{eq-D}
\end{equation}
is the magnetic flux current, $v$ is the vortex velocity, $\eta $ is the
Bardeen-Stephen drag (friction) coefficient \cite{BardSteph} for flux flow
and ${\cal A}$ is a numerical factor. Note that $D$ is proportional to the
electric field $E=(B\times v)/c$ in the sample. The form of the magnetic
flux current $D$ is chosen such that at $U=0$ and ${\cal A}=1$ the flux
velocity $v$ corresponds to the Bardeen-Stephen expression \cite{BardSteph}
for the flux flow: $v=v_{flow}=\phi _0j/c\eta $.

It has been already discussed \cite{Blatter-rev} that the strongly
non-linear Eq.~(\ref{eq-diff}) should obey a self-organized behavior. This
means that if a fluctuation $\delta U$ appears in the sample, it results in
a significant (exponential) local change of the flux current $D\propto \exp
(-\delta U/kT)$ which, in turn, leads to fast smearing out of the
fluctuation. In other words, $\left| \delta U\right| \simeq kT$ is the scale
of ''permitted'' variations of $U$. This has been proved experimentally by
direct measurements of $U$ using the Hall probe technique \cite{Abulafia1}.
We suggest a more general criterion of self-organization as follows.

The variation of the flux current density $D$ (or the same, of the electric
field $E$) within the sample can be written as: 
\begin{equation}
\delta D=\frac{\partial D}{\partial B}\delta B+\frac{\partial D}{\partial j}%
\delta j+\frac{\partial D}{\partial U}\delta U,  \label{eq-deltaD}
\end{equation}
where $B$, $j$ and $U$ are formally considered as three relaxing parameters,
though $U=U(B,j)$. If one of the three terms in the right-hand-side of Eq.~(%
\ref{eq-deltaD}) appears to be considerably greater by absolute value than
the other two terms, the corresponding parameter governs the relaxation
process, i.e., relaxes towards its mean value irrespective of what happens
with the other two. Obviously, this leads to a self-organization of the flux
diffusion process which implies that the three terms in the right-hand-side
of Eq.~(\ref{eq-deltaD}) tend to keep the same order: 
\begin{equation}
\frac{\partial D}{\partial B}\delta B\cong \frac{\partial D}{\partial j}%
\delta j\cong \frac{\partial D}{\partial U}\delta U.  \label{eq-order}
\end{equation}
The above condition can be considered as a mutual confinement for variations
of $B$, $j$ and $U$. Taking the expression for $D$ from Eq.~(\ref{eq-D}), we
get a limitation for $\delta U$: 
\begin{equation}
\frac{\delta U}{kT}\lesssim \max \left\{ \frac{\delta B}B,\frac{\delta j}j%
\right\} .  \label{eq-deltaU1}
\end{equation}
Note that the above estimation does not require the condition $U\gg kT$,
i.e., it should hold starting from the very early stages of flux creep.

If one of the three terms in Eq.~(\ref{eq-order}) is very small (or absent)
for any ''external'' reason, then the self-organization applies to the two
other ones. For instance, if $B$ is much greater than $H^{*}$ and thus $%
\delta B/B\ll 1$, we get: 
\begin{equation}
\frac{\delta U}{kT}\simeq \frac{\partial U}{\partial j}\frac{\delta j}{kT}%
\simeq \frac{U_c}{kT}\frac{\delta j}{j_c}<1,  \label{eq-deltaU2}
\end{equation}
where $U_c$ is the characteristic activation energy for $j\rightarrow 0$.
Since in general $U_c\gg kT$, we get $\delta j<\left( kT/U_c\right) j_c\ll
j_c$, as has already been discussed in Ref.~\cite{FGV}.

It is worth mentioning, however, that at the locations where $j=0$ or $B=0$
the variations of $U$ can exceed $kT$ significantly, as follows from Eq.~(%
\ref{eq-deltaU1}). The first of these two conditions ($j=0$) regularly holds
at the center of the sample, and we will comment on this point in the
following Section III. The second condition ($B=0$) holds at the lines where
vortices of different sign annihilate each other, or just at the edge of the
sample in the remanent state. We will devote special Section V for the
latter case.

Eqs.~(\ref{eq-deltaU1})-(\ref{eq-deltaU2}) prove the spatial constancy of $U$
throughout the sample with a $kT$ precision (see Eq.~(\ref{eq-const})). The
analytical results based on Eq.~(\ref{eq-const}) we will refer below as
''semi-analytical'' ones.

\section{One-dimensional creep equation}

In an infinite slab the current $j$ is related to $B$ by the Maxwell law: 
\begin{equation}
j=-\frac c{4\pi }\frac{\partial B}{\partial x},  \label{eq-Maxwell}
\end{equation}
if one uses a reference system $xyz$ where $B\parallel z$ and $j\parallel y$%
. For a platelet sample in a perpendicular external field, where the
in-plane field component $B_x$ appears, the relation between $j$ and $B$
becomes more complicated \cite{Brandt-rev,Zeldov}. Here we focus on the
one-dimensional creep problem where $B_x=0$, so $B_z=B$.

After substituting Eq.~(\ref{eq-Maxwell}) into Eq. (\ref{eq-diff}) one gets
the basic one-dimensional equation for flux motion: 
\begin{equation}
\frac{\partial B}{\partial t}=\frac \partial {\partial x}\left( {\cal A}%
\frac{\phi _0}{4\pi \eta }B\frac{\partial B}{\partial x}\exp (-U/kT)\right) .
\label{eq-basic}
\end{equation}
The numerical coefficient ${\cal A}$ depends on the creep mechanism and
should not necessarily be of order of unity \cite{ButLan} (see also
discussion in Ref.~\cite{FGV}). However, the low field measurements (i.e.,
in the single vortex pinning regime) in YBa$_{{\rm 2}}$Cu$_{{\rm 3}}$O$_{%
{\rm 7-x}}$ crystals \cite{Abulafia1}, where Eq.~(\ref{eq-basic}) was
experimentally studied by direct local measurements of $B$, $\partial
B/\partial x$ and $\partial B/\partial t$, revealed that ${\cal A}\simeq 1$.
Thus, at least in the case of single vortex pinning, Eq.~(\ref{eq-basic}) is
consistent with the equation for flux flow: 
\begin{equation}
\frac{\partial B}{\partial t}=\frac \partial {\partial x}\left( \frac{\phi _0%
}{4\pi \eta }B\frac{\partial B}{\partial x}\right) ,  \label{eq-flow}
\end{equation}
since the latter can be obtained from Eq.~(\ref{eq-basic}) at $U=0$ and $%
{\cal A}=1$.

The features of self-organized criticality in the solution of Eq.~(\ref
{eq-basic}) have been analyzed \cite{VFG} for the case of switching on of a
small additional field $\delta B$ on the background of $B\gg \delta B$
already present in the sample, and for a specific (logarithmic) dependence
of the activation energy on the current: $U=U_0\ln (j_0/j)$. In terms of the
energy distribution $U(x)$ across the sample the case considered in Ref.~ 
\cite{VFG} implies that in the beginning the energy was very large (or
infinite) in the whole sample, since $j=0$, then an area of small $U$
appeared at the edge (after switching on $\delta B$), and the propagation of
this ''fluctuation'' of the $U(x)$ profile was studied.

In contrast with Ref.~\cite{VFG} we consider below the instantaneous
switching on or removal of the whole external field $H$. A sort of
self-organization, i.e., establishing of a ''partial critical state'' \cite
{Beek-long} with $j(B)\propto j_c(B)$ has already been reported for this
case. We show below that this result follows from our more general approach
based on Eq.~(\ref{eq-const}) if $U=U(j/j_c(B))$, but for an arbitrary $%
U(B,j)$-dependence the partial critical state may not be established. Our
general results on self-organization of the flux creep do not depend on the
specific $U(B,j)$-dependence, but focus on the collective creep behavior $%
U(B,j)\propto B^\alpha j^{-\mu }$ as mostly relevant in HTSC. We will not
consider the time-dependent boundary conditions $H=H(t)$. Some results for
the latter case can be found in Refs. \cite{Bryksin,Gilchrist}.

The integration of Eq.~(\ref{eq-basic}) can be performed as follows.
Defining the magnetization as: 
\begin{equation}
m=\frac 1{2d}\int_{-d}^d\left( B-H\right) dx  \label{eq-m}
\end{equation}
and integrating Eq.~(\ref{eq-basic}) over $x$, we get: 
\begin{equation}
\frac{\partial m}{\partial t}={\cal A}\frac{\phi _0}{4\pi \eta d}H\left| 
\frac{\partial B}{\partial x}\right| _{x=\pm d}\exp (-U_{edge}/kT),
\label{eq-int1}
\end{equation}
where $U_{edge}=U(x=\pm d)$ is the activation energy at the edges of the
slab. For the case (discussed below) of straight field profiles: $\left|
\partial B/\partial x\right| \cong const$, and constancy of the activation
energy (the latter implies that $U_{edge}$ can be substituted by the mean
and almost constant $U$ over the sample, see Eq.~(\ref{eq-deltaU2})), one
can rewrite Eq.~(\ref{eq-int1}) in the form: 
\begin{equation}
\frac{\partial m}{\partial t}\cong {\cal A}\frac m\tau \exp (-U/kT),
\label{eq-int2}
\end{equation}
where $\tau =2\pi \eta d^2/\phi _0H$.

\subsection{Flux flow ($j>j_c$)}

After switching on the external field $H$, the flux flow starts and lasts
until the vortices fill the sample up to the critical profile $j_c$. Its
duration $\tau _{flow}$ can be easily estimated if one notices that already
during the flow regime the $B(x)$-profiles are almost straight, i.e., $%
\left| j\right| =(c/4\pi )\left| \partial B/\partial x\right| \simeq const$
(see dashed lines in Fig.~1a). Using the straightness of the field profiles,
one can estimate the magnetization $m$ as: 
\begin{eqnarray}
m(j) &\cong &H-\frac{cH^2}{8\pi jd}\qquad {\rm for}\quad j>j^{*},
\label{eq-m1} \\
m(j) &\cong &\frac{2\pi jd}c\qquad \qquad {\rm for}\quad j<j^{*},
\label{eq-m2}
\end{eqnarray}
where the current $j^{*}=cH/4\pi d$ shown as a grey solid line in Fig.~1a
discriminates between the incomplete and complete penetration of flux into
the sample. Note that we have chosen $H>H^{*}$, where $H^{*}=4\pi j_cd/c$ is
the field of full penetration. This means that $j_c<j^{*}$, i.e., by the
completion of the flux flow stage, flux penetrates the whole sample (see
Fig.~1a). Then, substituting these values into Eq.~(\ref{eq-int1}) and
solving it at $U=0$ and ${\cal A}=1$, we get: 
\begin{eqnarray}
j &\cong &j^{*}\sqrt{\frac \tau {2t}},\qquad \qquad \qquad t<\tau /2
\label{eq-tau1} \\
j &\cong &j^{*}\exp \left( \frac 12-\frac t\tau \right) ,\qquad t>\tau /2
\label{eq-tau2}
\end{eqnarray}
where $\tau /2=\pi \eta d^2/\phi _0H$ is the time of full penetration in the
flux flow regime. For $j_c<j^{*}$ the time of establishing of the critical
profile after switching on the external field is $\tau _{flow}\equiv
t(j_c)=(\tau /2)(1+2\ln j^{*}/j_c)\simeq \tau $ as follows from Eq.~(\ref
{eq-tau2}).

The crossover from the flux flow to flux creep process is well defined,
i.e., the critical profile $j=j_c$ is established at $t=\tau _{flow}$ almost
exactly throughout the whole sample (see Fig.~1a). More exactly, the
fluctuations of $U$ which appear in the whole slab at the crossover from
flow to creep $(j\cong j_c)$ are of order $kT$ according to Eq.~(\ref
{eq-deltaU2}), therefore $\delta j<(kT/U_c)j_c\ll j_c$. The only exclusion
is the very center of the slab, $x\cong 0$, where $j=0$ and $U$ shows
relatively strong variations, as we discussed in the previous Section.
However, this area is narrow and can be neglected when considering the
profiles $B(x)$ and magnetization $m$.

After the flux flow stage is completed, a much slower process of flux creep
starts, and various cases of $U(B,j)$ can be analyzed. First we consider the
simplest case where $U$ depends only on $j$.

\subsection{Creep at $j_c=const$, $U=U(j)$}

This case has been already studied in Refs.~\cite
{FGV,Beek-short,Beek-long,Schnack}. Here we analyze it as a test for our
numerical solution before consideration of more complicate models of $U(B,j)$%
. During the stage of flux creep $(j<j_c)$ (see Fig.~1a), the field profiles
are even more straight than during the flux flow stage \cite{FGV,Beek-short}%
, i.e., $\left| j\right| \cong const$, and, in turn, $U(j)\cong const$ (see
Fig.~1b). Note a very narrow increase of $\delta U$ at $x=0$, where $j=0$,
which is consistent with the comment at the end of Section II.

Since $U_{edge}$ is approximately equal to the mean $U$ over the sample, one
gets from Eqs.~(\ref{eq-int1}) and (\ref{eq-m2}), using also Eq.~(\ref
{eq-Maxwell}): 
\begin{equation}
\frac{dU}{dt}=\frac{dU}{dj}\frac{dj}{dm}\frac{dm}{dt}\cong \frac{{\cal A}}%
\tau j\left| \frac{dU}{dj}\right| \exp (-U/kT).  \label{eq-dUdt}
\end{equation}
This equation can be integrated numerically for any $U(j)$-dependence, and
also can be solved with a logarithmic accuracy \cite{GeshLar}, see Eq.~(\ref
{eq-Ulog}). The latter means that the real $U(j)$-dependence is substituted
by the tangent straight line with a slope $dU/dj$, as shown in Fig.~2, which
is reasonable since the relaxation slows down exponentially as $U$ grows.
Thus the system spends most of the relaxation time near the final point
where $U(j)$ and its tangent line almost coincide. Such an approximate
solution of Eq.~(\ref{eq-dUdt}) acquires the form of Eq.~(\ref{eq-Ulog})
with 
\begin{equation}
t_0=\frac{kT}{\left| dU/dj\right| }\frac{2\pi \eta d^2}{{\cal A}\phi _0jH}=%
\frac{\tau kT}{{\cal A}j\left| dU/dj\right| }.  \label{eq-t00}
\end{equation}
As becomes clear from Eq.~(\ref{eq-Ulog}) and Fig.~2, $t_0$ is the time
required to get from $U=-\infty $ (which corresponds to $t=-t_0$) to $U=0$
(which corresponds to $t=0$) along the non-physical part of the tangent
line, corresponding to negative $U$. Thus $t_0$ has no direct physical
meaning and should not be mixed with the characteristic duration $\tau
_{flow}\simeq \tau $ of flux flow, see Eqs.~(\ref{eq-tau1})-(\ref{eq-tau2}).

Eqs.~(\ref{eq-Ulog}) and (\ref{eq-t00}) provide a logarithmic approximation
for the time required for a system to reach the energy $U$. However, in
order to use Eq.~(\ref{eq-Ulog}) to describe the $U(t)$-dependence one
observes that $t_0$ is not actually a constant and depends on $U$ (or the
same, on $t$). This effect is not of great importance at $dU/dj\simeq const$%
, i.e., where $U$ is an almost linear function of $j$, but cannot be
neglected in the opposite case of strongly non-linear dependence of $U$ on $%
j $, where $dU/dj$ changes significantly.

Consider the case of the collective creep model 
\begin{equation}
U=U_c\left[ \left( j_c/j\right) ^\mu -1\right] ,  \label{eq-Ujcoll}
\end{equation}
which is an example of such a non-linear dependence. Here the exponent $\mu $
varies \cite{Blatter-rev} from $\mu =1/7$ (single vortex creep) to $\mu =5/2$
(small bundles). The straightforward solution of Eq.~(\ref{eq-dUdt}) then
gives: 
\begin{equation}
\mathop{\rm Ei}%
\left( \frac{U_c+U}{kT}\right) -%
\mathop{\rm Ei}%
\left( \frac{U_c}{kT}\right) ={\cal A}\mu \frac t\tau \exp \left( \frac{U_c}{%
kT}\right) ,  \label{eq-Utcoll}
\end{equation}
where $%
\mathop{\rm Ei}%
$ is the integral exponential function. The logarithmic approximation for
Eq.~(\ref{eq-Utcoll}) acquires the form: 
\begin{equation}
U=kT\ln \left( 1+{\cal A}\mu \frac t\tau (U+U_c)/kT\right) ,
\label{eq-Ulogcoll}
\end{equation}
which, if compared with Eq.~(\ref{eq-Ulog}), implies: 
\begin{equation}
t_0=\frac{\tau kT}{{\cal A}\mu (U+U_c)}.  \label{eq-t01}
\end{equation}
If during the creep process $j$ decreases down to $j_{\min }\ll j_c$, then
the energy increases up to $U_{\max }\gg U_c$, where $U_{\max }=U(j_{\min })$%
, and $t_0$ decreases down to $t_0^{\min }\cong \tau kT/{\cal A}\mu U_{\max
} $. Regularly, the latter estimation, $t_0\cong t_0^{\min }$, is
substituted into the logarithmic solution Eq.~(\ref{eq-Ulog}), and time
dependence of $t_0$ is neglected. This results in an almost linear
dependence of $U$ on $\ln (t)$ (see Fig.~3). Let us call such an
approximation (where $t_0$ is treated as a constant) as a ''pure''
logarithmic solution, whereas Eqs.~(\ref{eq-Ulogcoll})-(\ref{eq-t01})
provide a ''generalized'' one.

The straightforward solution and both generalized and pure logarithmic ones
are compared in Fig. 3. In the same figure we show an exact numerical
solution of Eq.~(\ref{eq-basic}) obtained without any assumptions on spatial
constancy of $U$. We used $\tau /{\cal A}\mu $ as a useful time constant in
Fig.~3. One observes that the generalized logarithmic solution works at all $%
t$, and together with the straightforward solution provide a perfect fit to
the exact one. On the other side, the pure logarithmic solution shows
significant deviations from the exact one, especially at short times. This
is consistent with Eq.~(\ref{eq-t01}), since there $U\lesssim U_c$, and, in
turn, $t_0\gg t_0^{\min }$. Particularly, the logarithmic solution misses
the characteristic maximum in the creep rate $dU/d\ln t$ which appears at $%
{\cal A}\mu t/\tau \cong 10$. For $\eta \simeq 10^{-5}\,{\rm g/cm\cdot }\sec 
$ \cite{Golos} (which implies that $T$ is well below $T_c$), ${\cal A}\simeq
1$, $\mu =1/7$ (single vortex creep) and $d=1\,{\rm mm}$, the maximum in $%
dU/d\ln t$ appears at $t^{*}\simeq 60/H\,\sec $, where $H$ is measured in $%
{\rm Oe}$. Thus the maximum or at least its tail can be resolved at
experimentally accessible times. Note that the numerical factor in the
latter estimation grows $\propto d^2$, and for some larger samples $t^{*}$
can be significantly greater. The position of such a maximum determined
experimentally can be used for determination of the parameters $\tau /{\cal A%
}$ and $\mu $.

The above theoretical results are compared in Fig.~4 with the experimental
data obtained on a large ($d\cong 1\,{\rm mm}$) YBa$_2$Cu$_3$O$_7$ crystal.
The relaxation of the magnetic moment was fitted by the direct numerical
solution of Eqs.~(\ref{eq-int2}) and (\ref{eq-Ujcoll}), where $\tau /{\cal A}
$, $j_c$, $\mu $ and $U_c$ were considered as independent fitting
parameters. Thus $U(t)$-dependence is found directly from the experiment and
is compared in Fig.~4 with the theoretical curve found from Eq.~(\ref
{eq-Utcoll}) with the same parameters. The experimental and theoretical
results almost coincide, and the characteristic maximum in $d(U/kT)/d\ln t$
at $t\cong 10\tau /{\cal A}\mu $ is very clear.

\subsection{Creep at $j_c=const$, $U=U_0(B/B_0)\left( j_c/j-1\right) $}

Consider this simplest model dependence of $U$ on $B$ and $j$, which can
mostly be analyzed analytically before a more complicated case of collective
creep. We have changed the notation $U_0$ instead of $U_c$, as in the
previous subsections, for reasons which are clarified below. Fig.~5
illustrates the numerical solution for $B(x)$ and $U(x)$ profiles in the
case. One observes that the general condition $U\cong const$ holds in this
case as well as in the previous one, where $U$ was independent of $B$. The
same narrow peak in $U$ is located at the center of the sample $x=0$, as
described in Section~II. Since $U$ depends only on the ratio $U_0/B_0$, we
can choose $B_0$ arbitrarily. It is most convenient to accept $B_0\equiv
H^{*}$. Then, using the Maxwell equation~(\ref{eq-Maxwell}) and the
condition $U\cong const$, one obtains the approximate expression for the
field profile in the sample: 
\begin{equation}
\frac xd\cong 1+\frac{B-H}{H^{*}}+\frac U{U_0}\ln \frac BH.
\label{eq-profile1}
\end{equation}
Denoting the field at the center of the sample as $B(0)$, which is found
from Eq.~(\ref{eq-profile1}) at $x=0$, one can rewrite the magnetic moment $%
m $ (see Eq.~(\ref{eq-m})), as: 
\begin{equation}
m=\frac 1d\int_{B(0)}^HxdB.  \label{eq-m-add}
\end{equation}
Then 
\[
\frac{\partial m}{\partial t}=\frac{\partial m}{\partial U}\frac{\partial U}{%
\partial t}=-\frac 1{U_0}\left[ \ln \frac H{B(0)}-H+B(0)\right] \frac{%
\partial U}{\partial t}. 
\]
On the other hand, as follows from Eq.~(\ref{eq-int1}), 
\begin{equation}
\frac{\partial m}{\partial t}=-\frac{{\cal A}}{4\tau }\left| \frac{\partial B%
}{\partial x}\right| _{x=\pm d}\exp (-U/kT),  \label{eq-dmdt-edge}
\end{equation}
and one gets an equation which determines the activation energy: 
\begin{equation}
\frac{dU}{dt}=\frac{{\cal A}U_0}\tau \frac h{2(h+u)\left[ h-b(0)-b(0)\ln
(h/b(0))\right] }\exp (-U/kT),  \label{eq-dUdt-mod0}
\end{equation}
where we denoted $h=H/H^{*}$, $b=B/H^{*}$, and $u=U/U_0$. This cumbersome
expression can be reduced if one uses the expansion over $\Delta =\left[
H-B(0)\right] /H$, which becomes exact at $\Delta \rightarrow 0$, but
actually holds with $10\%$ accuracy at worst for all $\Delta <1$, i.e., for
all $B(0)<H$. Then one gets: 
\begin{equation}
\frac{dU}{dt}\cong \frac{{\cal A}}\tau \left( U+U_c\right) \left( 1+\frac 23%
\Delta \right) \exp (-U/kT),  \label{eq-dUdt-mod}
\end{equation}
where $U_c=hU_0$. This result differs from Eq.~(\ref{eq-Ulogcoll}) only by
the correction factor $1+2\Delta /3$. This means that the results of the
previous subsection, where $U$ was field independent, apply here just by
accounting for the correction factor $1+2\Delta /3$, which in most cases is
not of great importance.

Eq.~(\ref{eq-dUdt-mod}) enables us to find the $U(t)$-dependence, and then,
using Eq.~(\ref{eq-profile1}) we get the field profile $B(x)$ as a function
of time, i.e., solve the creep problem completely. We call such an approach
as ''semi-analytical'' solution. Its results are compared in Fig.~5a with
the exact solution obtained by a direct numerical integration of Eq.~(\ref
{eq-basic}) with no assumptions on constancy of $U$. One observed that the
semi-analytical solution, being much less time-consuming (note that the
solution of Eq.~(\ref{eq-dUdt-mod}) is quite universal and has been already
obtained in the previous subsection), provides a perfect fit to the exact
description of the creep process.

\subsection{Creep at $j_c=const${\bf , }$U=U_0(B/B_0)^\protect\protect\alpha
\left( j_c/j\right) ^\protect\protect\mu $ (collective creep)}

This is the general dependence of $U$ on $B$ and $j$ in the collective creep
theory \cite{Blatter-rev} for $j\ll j_c$. The spatial constancy of $U$ holds
in this case as well as in previously considered cases. We have checked it
numerically for various $\alpha $ and $\mu $. The condition $U\cong const$
together with the Maxwell equation~(\ref{eq-Maxwell}) determines the field
profile: 
\begin{equation}
h^\nu -b^\nu \cong \frac \nu {u^{1/\mu }}\left( 1-\frac xd\right) ,
\label{eq-collprof}
\end{equation}
where we denote $\nu =1-\alpha /\mu $ and assume $B_0=H^{*}$, as in the
previous case. Here we take $\alpha \neq \mu $ (the case $\alpha =\mu $ is
almost identical to that considered in the previous subsection). The field $%
b(0)$ is determined, as follows from Eq.~(\ref{eq-collprof}), by $h^\nu
-b(0)^\nu =\nu /u^{1/\mu }$. The magnetic moment can be calculated using
Eq.~(\ref{eq-m-add}): 
\begin{equation}
m=H^{*}\left[ h-\frac{u^{1/\mu }}{\nu +1}\left( h^{\nu +1}-b(0)^{\nu
+1}\right) \right] ,  \label{eq-collm}
\end{equation}
and, using Eq.~(\ref{eq-dmdt-edge}), one gets: 
\begin{equation}
\frac{dU}{dt}=\frac{{\cal A}\mu (\nu +1)}{2\tau h^{\nu -1}u^{2/\mu }}\left[
h^{\nu +1}-b(0)\left( h^\nu +u^{-1/\mu }\right) \right] ^{-1}U\exp (-U/kT).
\label{eq-dUdt-coll0}
\end{equation}
This expression, being a bit cumbersome, can be reduced using the expansion
over $\Delta =(H-B(0))/H$, which, as in the previous case, works with
reasonable accuracy (better than $10\%$) at all $B(0)\lesssim H$: 
\begin{equation}
\frac{dU}{dt}=\frac{{\cal A}\mu }\tau U\left[ 1+\frac{2\alpha }{3\mu }\Delta %
\right] \exp (-U/kT).  \label{eq-dUdt-coll}
\end{equation}
This result is very similar to Eq.~(\ref{eq-dUdt-mod}). The absence of $U_c$
in Eq.~(\ref{eq-dUdt-coll}) corresponds to the absence of term $-1$ in the
dependence of the activation energy $U$ on $j$ in this model (compared with
the two previous cases). The factor in the square brackets in Eq.~(\ref
{eq-dUdt-coll}) describes the effective renormalization of $U$ in the
pre-exponential factor resulting from the dependence of $U$ on $B$. If $%
\alpha =0$, which means that $U$ is independent of $B$, then the
renormalization disappears. The same happens at $B(0)\rightarrow H$. At $%
\alpha =\mu $ the correction factor reduces to $1+2\Delta /3$, which is
consistent with the previous case, where $\alpha =\mu =1$.

In Fig.~6 we compare the direct numerical solution of Eq.~(\ref{eq-basic})
with the semi-analytical one determined by Eqs.~(\ref{eq-collprof})-(\ref
{eq-collm}). Fig.~6a shows the numerical (exact) $B(x)$ profiles compared
with Eq.~(\ref{eq-collprof}), and Fig.~6b shows $m(\ln t),$ obtained
numerically from Eq.~(\ref{eq-basic}) and semi-analytically from Eqs.~(\ref
{eq-collprof})-(\ref{eq-dUdt-coll}). The quality of the semi-analytical
approach is perfect in this case as well as in the previous one.

In the most general case: 
\begin{equation}
U=U_0(B/B_0)^\alpha \left[ \left( j_c/j\right) ^\mu -1\right]
\label{eq-general}
\end{equation}
one gets an expression which naturally conforms to Eqs.\ (\ref{eq-dUdt-mod})
and (\ref{eq-dUdt-coll}): 
\begin{equation}
\frac{dU}{dt}=\frac{{\cal A}\mu }\tau \left( U+U_c\right) \left[ 1+\frac{%
2\alpha }{3\mu }\Delta \right] \exp (-U/kT),  \label{eq-dUdt-general}
\end{equation}
where, as above, $U_c=hU_0$. We skip the cumbersome derivation of the last
expression, which requires expansion over $\Delta $ starting from the
equation for the field profile $B(x)$. The generalized logarithmic solution
of this equation acquires the form 
\begin{equation}
U=kT\ln \left( 1+\frac{{\cal A}\phi _0H\mu (U+U_c)\left[ 1+(2\alpha /3\mu
)\Delta \right] }{2\pi \eta d^2kT}t\right) ,  \label{eq-Ulog-general}
\end{equation}
which coincides with Eq.~(\ref{eq-Ulog}) if 
\begin{equation}
t_0=\frac{2\pi \eta d^2kT}{{\cal A}\phi _0H\mu (U+U_c)\left[ 1+(2\alpha
/3\mu )\Delta \right] }=\frac{H^{*}}Ht_H,  \label{eq-t0-general}
\end{equation}
where we introduced $t_H\equiv t_0(H=H^{*})$. Note that $t_H$ is almost
field independent, since $H$ enters $t_H$ only via $\Delta $. Note that both 
$t_0$ and $t_H$ depend on time via $U$ and the correction term (in square
brackets).

\subsection{Creep at $j_c=j_c(B)\neq const$}

Above, we have considered only $j_c=const$. However, the field dependence of
the critical current, $j_c=j_c(B)$, does not violate the general condition,
Eq.~(\ref{eq-const}), of self-organization of flux creep. As a direct
consequence of this condition it is worth mentioning the following: If the
dependence of $U$ on $j$ and $B$ has the form: $U\propto f(j_c(B)/j)$, where 
$f$ is an arbitrary function, then the spatial constancy of $U$ results in
establishing of a ''partial'' critical state \cite{Beek-long} with $j\propto
j_c(B)$. For instance, if the critical current obeys the Kim dependence $%
j_c(B)=j_0B_0/(B_0+B)$, then the field profile $B(x)$ during the relaxation
should be determined by the condition $j=pj_0B_0/(B_0+B)$ with $0<p<1$.

However, for more complicated dependencies of $U$ on $B$ and $j$ this is not
the case, and the profiles $B(x)$ can differ significantly from that in the
critical state. In the next Section we consider an example of such a
behavior.

\section{Semi-analytical solutions for anomalous magnetization (fishtail).}

The equation (\ref{eq-dUdt-general}) and its reduced forms (see Eqs.~(\ref
{eq-dUdt-mod}) and (\ref{eq-dUdt-coll})), which are just ordinary
differential equations, present the method of semi-analytical integration of
the equation for flux motion (see Eq.~(\ref{eq-basic})), for the case of
collective creep, where the dependence of $U$ on $B$ and $j$ is described by
Eq.~(\ref{eq-general}), or by its reduced versions. Of course an analogous
solution can be found for any $U(B,j)$-dependence, not only for that
described by Eq.~(\ref{eq-general}). This semi-analytical approach provides
a good fit to the exact solution, obtained by numerical integration of Eq.~(%
\ref{eq-basic}), as one can see in Fig.~6. The correction factor $1+(2\alpha
/3\mu )\Delta $ can be neglected except for short times $t\gtrsim \tau /%
{\cal A}\mu $.

The semi-analytical solutions can be applied for the description of an
anomalous magnetization, coined a ''fishtail'', found in clean high-T$_{{\rm %
c}}$ superconductors \cite{fisht,Abulafia2,Giller}. Note that $j_c$ enters
Eq.~(\ref{eq-dUdt-general}) only via the correction factor which is
negligible in most cases, especially at high fields $H\gg H^{*}$ where $%
\Delta \ll 1$. Thus the solution $U(t)$ of Eq.~(\ref{eq-dUdt-general}) is
determined by the current exponent $\mu $ (and not by the field one, $\alpha 
$), by the characteristic energy $U_c$, and by $\tau $ (which in turn
depends on $d$, $\eta $ and $H$). If one measures the magnetization current $%
j$ at the edge of the sample, where $U(B,j)=U(H,j)$ as a function of $H$,
keeping the time window $t$ of the experiment constant for each $H$ (this is
the case for most studies of fishtails), then $U$ along the measured line $%
M(H)$ or $j(H)$ can be written, as follows from Eq.~(\ref{eq-Ulog-general}),
as 
\begin{equation}
U(H,t)/kT-\ln (U+U_c)\cong \ln H+\ln t+\ln \left( \frac{{\cal A}\phi _0\mu %
\left[ 1+(2\alpha /3\mu )\Delta \right] }{2\pi \eta d^2kT}\right) ,
\label{eq-UkTlnH}
\end{equation}
where the last term in Eq.~(\ref{eq-UkTlnH}) almost does not depend on $H$
and $t$. This means that the magnetization curve is determined by: 
\begin{equation}
\frac{dU}{kT}-\frac{dU}{U+U_c}=\frac{dH}H.  \label{eq-difference}
\end{equation}
In Fig.~7 we present the results of our semi-analytical approach to the
problem of dynamic fishtail formation taking $j_c(B)=j_0B_c/(B_c+B)$ (Kim
model) and collective creep with 
\begin{equation}
U=U_0(B/H^{*})^\alpha \left[ \left( \frac{j_c(B)}j\right) ^\mu -1\right]
\left( \frac{B_c+B}B\right) ^\mu .  \label{eq-Usimul}
\end{equation}
The last factor in this equation is added to cancel the dependence of $j_c$
on $B$ and keep the general collective creep condition: $U\propto B^\alpha
j^{-\mu }$ for $j\ll j_c$. At each $H$ we find the energy $U$ down to where
the system relaxes during the ''experimental'' time window $t$, and then,
using this $U$, we determine the corresponding $j_{x=\pm d}$ according to
Eq.~(\ref{eq-Usimul}). The results show a clear fishtail due to fast
relaxation at low fields (see Fig.~7).

Note that Eq.~(\ref{eq-Usimul}) provides an example of the case where the
field profiles $B(x)$ are significantly different from the critical one at $%
j=j_c$, and a ''partial critical state'' \cite{Beek-long} is not established.

Since $dU=(\partial U/\partial H)dH+(\partial U/\partial j)dj$, one obtains
using Eq.~(\ref{eq-difference}) that the magnetization curve (fishtail) is
determined by the condition: 
\begin{equation}
\frac{dj}{dH}=-\left( \frac{\partial U}{\partial H}-\frac{kT}H\frac{U+U_c}{%
U+U_c-kT}\right) \left( \frac{\partial U}{\partial j}\right) ^{-1}.
\label{eq-djdH1}
\end{equation}
For the case of collective creep, where $U\propto H^\alpha j^{-\mu }$, we
have $\partial U/\partial H=(\alpha /H)U$ and $\partial U/\partial j=-(\mu
/j)U$. Then, taking into account that $U+U_c\gg kT$, we get 
\begin{equation}
\frac{dj}{dH}\cong \frac \alpha \mu \frac jH\left( 1-\frac 1\alpha \frac{kT}U%
\right) .  \label{eq-djdH2}
\end{equation}
The peak of the fishtail, where $j(H)$ reaches maximum, corresponds to $%
U\cong kT/\alpha $, as follows from Eq.~(\ref{eq-djdH2}). This implies that $%
j$ increases as a function of $H$ until it almost reaches the $j_c(H)$
curve. Far below $j_c$, where $kT/\alpha U\ll 1$, one gets from Eq.~(\ref
{eq-djdH2}) that $j\propto H^{\alpha /\mu }$.

We see from Eqs.~(\ref{eq-UkTlnH}) and (\ref{eq-difference}) that $U$
changes along the magnetization curve obtained at a fixed time window $t$.
However, one can measure $j_{Ht}(H)$ keeping the product $Ht$ as constant,
which, according to Eq.~(\ref{eq-Ulog-general}), should result in a constant 
$U$ along the magnetization curve (neglecting the correction factor $%
1+(2\alpha /3\mu )\Delta $). The difference $j_{Ht}(H)-j(H)$, where $j(H)$
is taken at $t=const$, is determined by: 
\begin{equation}
\frac{d(j_{Ht}-j)}{dH}\cong -\frac{kT}{H\left( \partial U/\partial j\right) }%
,  \label{eq-dj}
\end{equation}
which provides a tool for independent analysis of $U(j)$-curve.

Above in this Section we have considered the exponents $\alpha $ and $\mu $
to be constants. However, different regions in the $j-H$ diagram correspond
to different relaxation regimes, such as single vortex creep, small and
large bundle creep, etc. (see \cite{Blatter-rev,YSM-rev,Brandt-rev}). The
energy scale $U_0$, as well as the exponents $\alpha $ and $\mu $ may vary
significantly from one region of $j-H$ to another. As one observes from Eq.~(%
\ref{eq-Ulog-general}), the crucial exponent of the above two is $\mu $. Its
rapid change at the boundary between the creep regions from $\mu _1$ to $\mu
_2$ is equivalent to a change of $H$ by factor $\mu _2/\mu _1$. As follows
from Eq.~(\ref{eq-UkTlnH}), this results in a change of $\simeq \ln (\mu
_2/\mu _1)$ in $U/kT$ at the boundary between two creep regions. Thus $U$
does not change much at the crossover from one pinning regime to another.
However, $j$ (and, in turn, $M$) can be changed significantly at such a
boundary, since for different relaxation laws (different $U_c$, $\alpha $
and $\mu $) the same $U$ is reached at significantly different $j$. As $H$
increases, the growing vortex bundles lead to increase of characteristic
energies $U_c$, thus one should expect a step-like increase of $j$ when
crossing the boundaries {\it single vortex pinning} $\rightarrow $ {\it %
small bundles} $\rightarrow $ {\it large bundles}.

If one measures the exponent $\mu $ along the magnetization curve (see, for
instance, Ref.~\cite{Abulafia2}), then a curve of constant $U$ in the $H-j$
diagram can be plotted using rather $t\propto (\mu H)^{-1}$ instead of $%
t\propto H^{-1}$, as was suggested above.

\section{Relaxation in the remanent state and annihilation lines}

A particular and very interesting case, where the discussed above
self-organization of the flux motion should be modified significantly, is
relaxation in the presence of annihilation lines $B=0$. The vortices and
antivortices approach the annihilation line from different sides and
annihilate each other. The arguments of Section II for the constancy of $U$
are not valid in this case, at least in the vicinity of the annihilation
lines (see comment at the end of Section II). Therefore, this case should be
studied separately.

Consider the simplest situation of remanent relaxation, where the field has
been ramped up and then instantaneously removed, so $B=0$ at the edges $%
x=\pm d$ of the slab. There are no antivortices in this case, since the
annihilation line coincides with the edge of the sample.

The description of the flux motion in this case using Eq.~(\ref{eq-diff})
looks self-contradictory since at the sample edge $B=0$, whereas the
magnetic flux current $D=Bv$ is finite at the edges and, moreover, obviously
should reach there its maximal value over the sample. However, the
contradiction is void provided the field vanishes at the sample edge as $%
B\propto \sqrt{d-x}$, i.e., proportional to the square root of the distance
to the edge (see Fig.~8). At the same time the vortex velocity diverges at
the sample edge as $v\propto \partial B/\partial x\propto 1/\sqrt{d-x}$.
This divergency is removed by an appropriate cut-off for $d-x$, which we
discuss later in this Section, but inevitably leads to the appearance of the
flux-flow region near the edge or, most generally, near the annihilation
line. However $D\propto B(\partial B/\partial x)$ remains finite and
continuous with no singularity at the edge. This is confirmed by direct
computer simulations of the relaxation in the remanent state (see Fig.~8).

Let us estimate the coefficient $k$ in the square-root dependence $%
B_{edge}\cong k\sqrt{d-x}$ near the sample edge, which should include the
flux flow region $U=0$. The magnetic flux current reaches at $x=d$ its
maximum over the sample: $D_{x=d}=(\phi _0/4\pi \eta )\left[ B\,\partial
B/\partial x\right] _{x\rightarrow d}=\phi _0k^2/8\pi \eta $, but remains of
the same order as the mean flux current $\left\langle D\right\rangle $ over
the sample: $D_{x=d}={\cal C}\left\langle D\right\rangle $, where ${\cal C}%
\gtrsim 1$ is a numerical factor. Estimating $\left\langle D\right\rangle
\simeq ({\cal A}\phi _0/4\pi \eta )\left\langle B\right\rangle \left\langle
\partial B/\partial x\right\rangle \exp (-\left\langle U\right\rangle /kT)$, 
$\left\langle B\right\rangle \simeq B(0)/2$, $\left\langle \partial
B/\partial x\right\rangle \simeq B(0)/d$, one gets $k\simeq B(0)\sqrt{{\cal %
CA}\exp (-\left\langle U\right\rangle /kT)/d}$. Note that the above
estimation is based on the constancy of $U$, i.e., $U\cong \left\langle
U\right\rangle $ throughout the sample except the edge flux-flow regions,
which we assume to be small. Thus we get: 
\begin{equation}
B_{edge}(x)\cong B(0)\sqrt{{\cal CA}\exp (-\left\langle U\right\rangle /kT)%
\frac{d-x}d}.  \label{eq-Bedge}
\end{equation}

A natural cut-off for the area of applicability of Eq.~(\ref{eq-Bedge}) is: $%
d-x>\lambda $, otherwise the surface effects such as Bean-Livingston
interaction with the surface \cite{BL} should be accounted for. There are
additional restrictions: (i) the current cannot exceed the depairing one: $%
j=(c/4\pi )\partial B/\partial x<j_d$; and (ii) the intervortex distance $%
a\simeq \sqrt{\phi _0/B(x)}$ should not exceed the distance to the surface $%
d-x$ at any $x$. It can be easily confirmed that the condition $d-x>\lambda $
is stronger than the other two at most reasonable values of $B(0)$ and $d$.

Let us call the region $\tilde x<\left| x\right| <d$ near the sample edges,
where the activation energy grows from $U=0$ at the very edge (flux-flow
region) up to $U(\tilde x)\cong \left\langle U\right\rangle $, as the area
of ''annihilation dominated'' organization of flux creep. Its width $d-%
\tilde x$ can be estimated as follows: We substitute Eq.~(\ref{eq-Bedge})
into the collective creep formula for $U(B,j)$, see Eq.~(\ref{eq-general}),
and find $\tilde x$ where $U$ reaches its mean value $\left\langle
U\right\rangle $. This is of course a crude approximation, since Eq.~(\ref
{eq-Bedge}) is valid in the flux-flow region only, and for the whole
''annihilation dominated'' region it provides an underestimation for $B$
and, in turn, $j$. After straightforward calculations we get: 
\begin{equation}
\frac{d-\tilde x}d\simeq \left( \frac{\left\langle U\right\rangle }{U_0}%
\right) ^{\frac 2{\alpha +\mu }}\exp \left( \frac{\alpha -\mu }{\alpha +\mu }%
\frac{\left\langle U\right\rangle }{kT}\right) .  \label{eq-width}
\end{equation}

The above result implies that the width $d-\tilde x$ of the ''annihilation
dominated'' region is crucially dependent upon the relationship between the
exponents $\alpha $ and $\mu $. For $\mu >\alpha $ and $\left\langle
U\right\rangle \gg kT$ this region appears to be exponentially small, i.e., $%
\tilde x\cong d$. Computer simulations show a step-like increase of $U$ at
the edge to the value comparable with $\left\langle U\right\rangle $, and
then $U$ grows smoothly and slowly towards the center of the sample (see
Fig.~9a). Though $\delta U$ appears to be significantly greater than for the
case of finite $H$, discussed in previous Sections, even here $U$ does not
vary significantly: $\delta U\lesssim 4kT$ in the whole sample, excluding
the sharp step at the edge. For the opposite case, $\mu <\alpha $, one finds
from Eq.~(\ref{eq-width}) the unphysical result that $(d-\tilde x)/d$ is
exponentially large though, of course, $(d-\tilde x)/d<1$ anyway. This
implies that our assumption about the spatial constancy of $U(x)\cong
\left\langle U\right\rangle $ throughout almost the whole sample (except
small edge regions) is self-contradictory in this case. Thus for $\mu
<\alpha $ the effect of the annihilation line spreads over the whole sample,
and there is not any evidence of constancy of $U$. This is confirmed by
numerical simulations (see Fig.~9b). The boundary case, where $\mu =\alpha $%
, is illustrated in Fig.~9c.

\section{Conclusion}

We considered the generalization of the logarithmic solution Eq.~(\ref
{eq-Ulog}) for flux creep at different dependencies of the activation energy 
$U$ on field $B$ and current $j$ and confirmed it by numerical analysis. The
general condition which governs the relaxation is $U(x)\cong const$
throughout the sample, and this result holds at any particular $U(B,j)$%
-dependence. This results from a {\em self-organization} of flux creep in
the undercritical state $j<j_c$, which implies that the influence of all the
creep parameters, $B$, $j$ and $U$, on the relaxation rate should be of the
same order of magnitude. This self-organization should not be mixed with the 
{\em self-organized criticality} of flux motion at $j\cong j_c$.

For $U$ independent of $B$, i.e., $U=U(j)$, we restore the known result of
straightness ($j\simeq const$) of the field profiles throughout the sample.
For the case where $U$ essentially depends on $B$ the condition of spatial
constancy of $U(B,j)$ determines the one-parameter family of field profiles $%
B_U(x)$ and enables us to find a semi-analytical solution for $U(t)$ and, in
turn, for time evolution of the field profiles $B_U(x)$, i.e., to solve the
creep problem completely. Such a semi-analytical solution provides a perfect
fit to the exact numerical solution (obtained without any assumptions on
constancy of $U$) and appears to be quite useful for the description of the
dynamic development of anomalous magnetization (fishtail) due to fast
relaxation rates at low fields.

The effect of the annihilation lines $B=0$ on the self-organization of the
collective creep where $U\propto B^\alpha j^{-\mu }$ (see Eq.~(\ref
{eq-general})) is crucially dependent on the relationship between $\alpha $
and $\mu $. At $\alpha <\mu $ the effect is just an increase of variation $%
\delta U$ over the sample, with a step-like vanishing of $U$ in the very
narrow regions of flux flow in the vicinity of an annihilation line.
However, for $U\gg kT$ we still get $\delta U\ll U$, i.e., the condition $%
U\cong const$ holds qualitatively in this case. At $\alpha >\mu $ the above
condition no longer holds, and the presence of an annihilation line destroys
the self-organization in the whole sample irrespective of it size.

{\em Acknowledgments.}

The authors are grateful to V. M. Vinokur for critical reading of the
manuscript and valuable discussions. L.~B. acknowledges support from the
Israel Academy of Sciences and from the German-Israeli Foundation (GIF).

\begin{center}
\vspace{1.0in}{\sc Figure captions}
\end{center}

\begin{enumerate}
\item[Fig. 1]  a) The field profiles $B(x)$ for flux flow (dashed lines) and
creep (squares) for $U=U_0\left[ \left( j_c/j\right) -1\right] $ with $%
H^{*}=H/2$. The critical state ($j=j_c$) and the full penetration state ($%
j=j^{*}$) are shown in solid black and grey lines, respectively. Note an
almost exact formation of a critical state at the crossover flow$\rightarrow 
$creep.\newline
b) Spatial dependence of $U$ at different times $t/\tau _{flow}$. Note
almost spatial constancy of $U$ except narrow regions $x\cong 0$.

\item[Fig. 2]  Relaxation of $j$ can be imagined as a motion of a dot along
the $U(j)$-curve (solid). For the logarithmic approximation $U(j)$ is
substituted by its tangent line (dashed). Time $t_0$ corresponds to the
''motion'' along the negative part ($U<0$) of the line, whereas $\tau
_{flow} $ corresponds to the ''motion'' from $j=\infty $ down to $j=j_c$
along $U=0$.

\item[Fig. 3]  Comparison of ''pure'' logarithmic (triangles),
straightforward (circles) and generalized logarithmic (solid line) solutions
of Eq.~(\ref{eq-dUdt}), which were derived under assumption $U(x)=const$,
together with the ''exact'' numerical solution (squares) of Eq.~(\ref
{eq-basic}) for $U(j)=U_c\left[ \left( j_c/j\right) ^\mu -1\right] $. Open
and filled symbols correspond to $U/kT$ and $d(U/kT)/d\ln t$, respectively.
Note that all the solutions except the pure logarithmic one show a maximum
in $dU/d\ln t$ at ${\cal A}\mu t/\tau \gtrsim 1$.

\item[Fig. 4]  Experimentally obtained relaxation rate $d(U/kT)/d\ln t$
(triangles) and normalized magnetic moment $M/M_c$, which is equal to $j/j_c$
(circles) vs. $t$ and their fit by the generalized logarithmic solution at
the same values of parameters: ${\cal A}/\tau =0.03$ sec$^{-1}$, $%
U_c/kT=12.62$ and $\mu =2.03$.

\item[Fig. 5]  a) The field profiles $B(x)$ for flux creep at $%
U=U_0(B/H^{*})(j_c/j-1)$ with $H^{*}=H/2$ found by numerical solution of
Eq.~(\ref{eq-basic}) (squares) and by the semi-analytical approach (solid
lines).\newline
b) $U(x)$ found for the same $U(B,j)$-dependence from the numerical solution
of Eq.~(\ref{eq-basic}).

\item[Fig. 6]  a) The same as in Fig.~5a for the collective creep dependence 
$U=U_0(B/H^{*})^\alpha (j_c/j)^\mu $ with $\alpha =1$, $\mu =2$ and $%
H^{*}=H/2$.\newline
b) Magnetization for the same $U(B,j)$-dependence found from the numerical
solution of Eq.~(\ref{eq-basic}) (circles) and by semi-analytical approach
(solid line). For $m/H<0.1$ the circles and the line completely coincide.

\item[Fig. 7]  Dynamic development of anomalous magnetization (fishtail)
found by the semi-analytical solution of the Kim model (see Eq.~(\ref
{eq-Usimul})) with $B_c=2H^{*},$ $\alpha =1$ and $\mu =2$. Relaxation starts
at $t=0$ from $j_c(H)=j_0B_c/\left( B_c+H\right) $ shown as dashed line. Due
to faster relaxation at small $H$ an anomalous magnetization develops at $%
j\ll j_c$. Circles and solid lines correspond to the direct numerical and
semi-analytical solutions, respectively, for $\ln (t/\tau _{flow})=2.8$, $%
7.4 $ and $14.3$.

\item[Fig. 8]  The magnetic induction $B$ (squares), vortex velocity $v$
(circles) and the magnetic current $D=Bv$ (triangles) vs. $d-x$ found
numerically from Eq.~(\ref{eq-flow}) near the sample edge $x=d$. The fits
are: $\sqrt{d-x}$ for $B$ and $1/\sqrt{d-x}$ for $v$. Note that $D$ shows no
peculiarity at $x=d$.

\item[Fig. 9]  The activation energy $U/kT$ in the remanent state: a) $%
\alpha =0.1$, $\mu =1$; b) $\alpha =1.5$, $\mu =0.5$ c) $\alpha =1$, $\mu =1$%
.
\end{enumerate}

\end{document}